\documentclass[sigconf]{acmart}

\AtBeginDocument{%
  \providecommand\BibTeX{{%
    \normalfont B\kern-0.5em{\scshape i\kern-0.25em b}\kern-0.8em\TeX}}}

\usepackage{subcaption}
\begin{document}

\title{Foodbot: A Goal-Oriented Just-in-Time Healthy Eating Interventions Chatbot}


\author{Philips Kokoh Prasetyo}
\email{pprasetyo@smu.edu.sg}
\affiliation{%
  \institution{Singapore Management University}
}

\author{Palakorn Achananuparp}
\email{palakorna@smu.edu.sg}
\affiliation{%
  \institution{Singapore Management University}
}

\author{Ee-Peng Lim}
\email{eplim@smu.edu.sg}
\affiliation{%
  \institution{Singapore Management University}
}


\newcommand\todo[1]{\textcolor{red}{#1}}
\begin{abstract}
  
Recent research has identified a few design flaws in popular mobile health (mHealth) applications for promoting healthy eating lifestyle, such as mobile food journals. These include tediousness of manual food logging, inadequate food database coverage, and a lack of healthy dietary goal setting. To address these issues, we present Foodbot, a chatbot-based mHealth application for goal-oriented just-in-time (JIT) healthy eating interventions. Powered by a large-scale food knowledge graph, Foodbot utilizes automatic speech recognition and mobile messaging interface to record food intake. Moreover, Foodbot allows users to set goals and guides their behavior toward the goals via JIT notification prompts, interactive dialogues, and personalized recommendation. 
 Altogether, the Foodbot framework demonstrates the use of open-source data, tools, and platforms to build a practical mHealth solution for supporting healthy eating lifestyle in the general population.
  
\end{abstract}




\maketitle

\section{Introduction}
\label{sec:motivation}


\begin{figure}[tp]
\centering
\begin{subfigure}[b]{0.235\textwidth}
\centering
\includegraphics[width=1\linewidth]{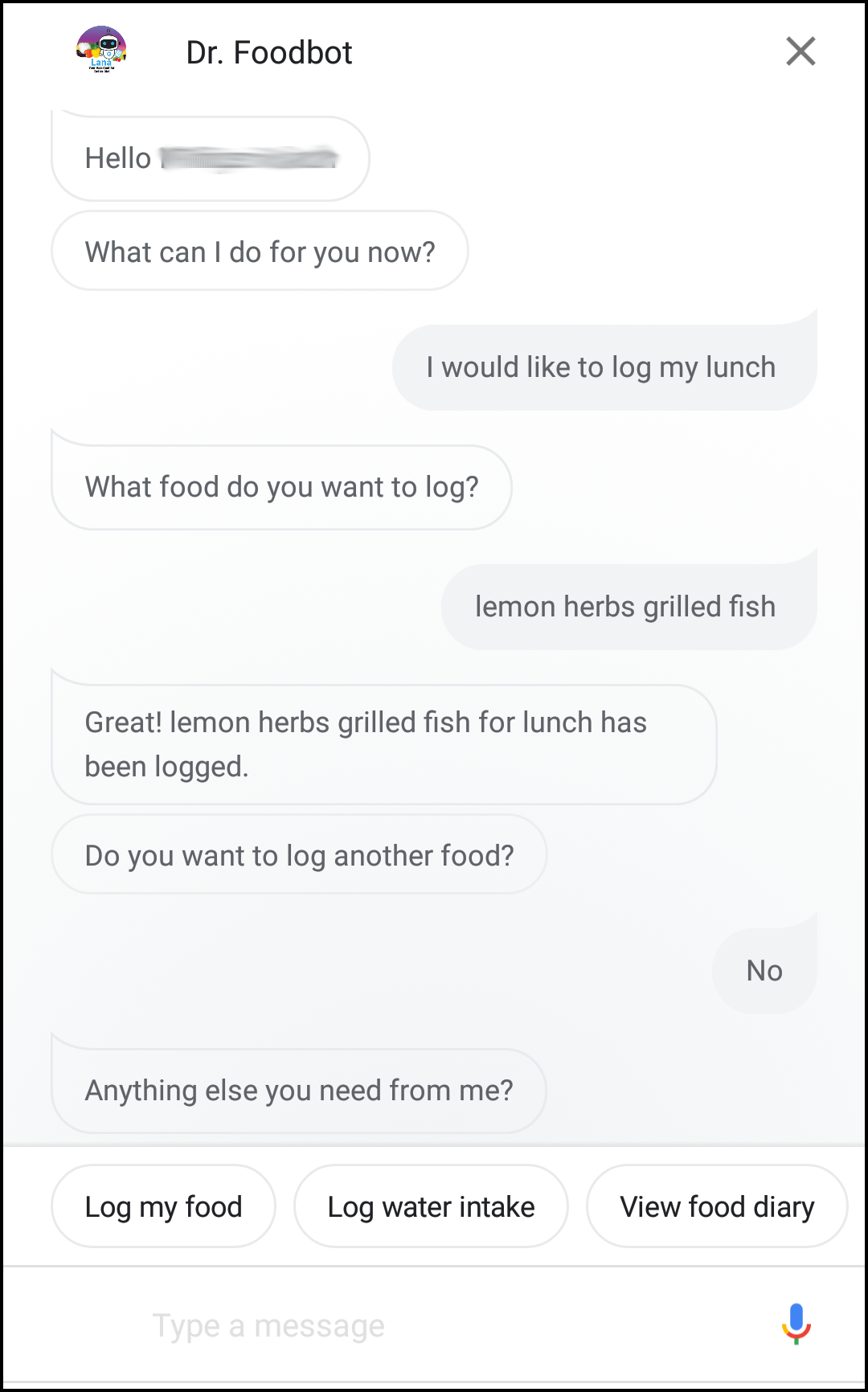} 
\caption{Food Logging}
\label{fig:screenshot_logging}
\end{subfigure}
\begin{subfigure}[b]{0.235\textwidth}
\centering
\includegraphics[width=1\linewidth]{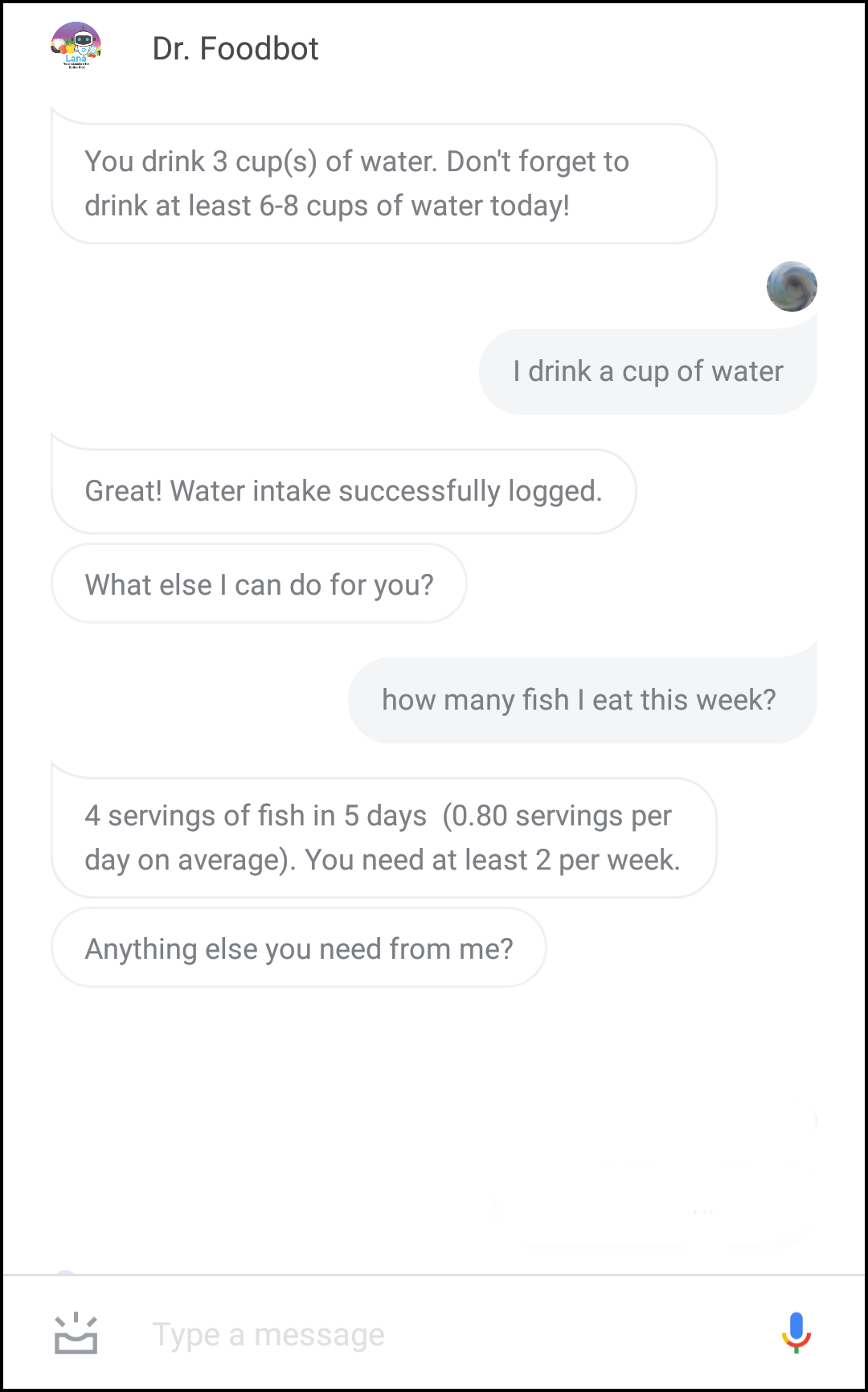}
\caption{Goal Tracking}
\label{fig:screenshot_intervention}
\end{subfigure}
\caption{Dialogue Turns in Foodbot}
\label{fig:screenshot_foodbot}
\end{figure}

\textbf{Motivation.}  Our health begins with what we eat. Maintaining healthy eating habit is one of the key factors in reducing the risk of chronic preventable diseases, such as coronary vascular diseases and type-2 diabetes, and increasing life expectancy. This generally involves consuming a healthful and balanced diet and practicing portion control on a regular basis. Nevertheless, a lack of motivation, poor self-regulation, and personal biases tend to prevent people from successfully developing and maintaining healthy eating behavior.





Subsequently, researchers have looked into mHealth and mobile sensing technologies as a cost-effective way to facilitating healthy eating lifestyle at a population scale. A popular type of mHealth applications (apps) for dietary self-tracking is mobile food journal, such as MyFitnessPal, 
in which users manually record their food entries by selecting items from a food database. However, previous studies \cite{Cordeiro2015BarriersANN,Achananuparp2018DoesJEH} recently identified design flaws in mobile food journal apps which lead to unintended and undesirable user behaviors.


Specifically, many users were frustrated by the tedious nature of manual data entry effort~\cite{Cordeiro2015BarriersANN}. 
Next, most food databases used by mobile food journal apps consist of a combination of commercially available and user-contributed food and nutritional records. Even so, many users were not able to find many food items, such as those from local restaurants \cite{Cordeiro2015BarriersANN}, due to poor data coverage. Lastly, unexpected lapses in healthy eating habits of mobile food journal users \cite{Achananuparp2018DoesJEH} suggested that mobile food journal apps tend to overly emphasize caloric and weight goals, whereas consumption of a healthful and balanced diet was not used as goals. Consequently, these design flaws limit the practicality and effectiveness of mHealth technologies in supporting healthy eating lifestyle.

\begin{figure*}[tp]
  \centering
  \includegraphics[width=0.78\textwidth]{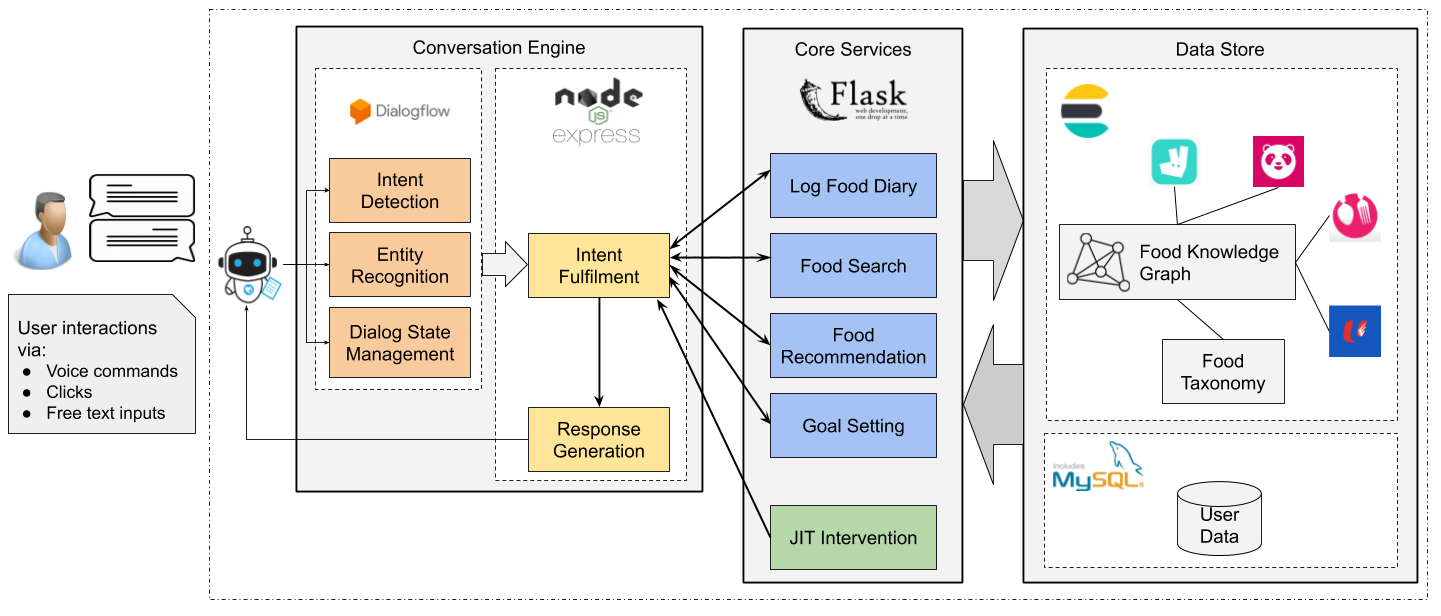}
  \caption{System Architecture}
  \label{fig:system}
\end{figure*}

\textbf{Objectives.} To address the aforementioned issues, we present Foodbot, a chatbot-based mHealth application for goal-oriented just-in-time healthy eating interventions.
The main contribution of Foodbot is in its open-source design incorporating: (1) a natural-language user interface; (2) a large-scale food knowledge graph; and (3) goal-setting and JIT intervention and recommendation, to overcome the burden of manual dietary self-tracking and provide users with a personalized guidance toward a healthy eating lifestyle. Its behavior change design follows common techniques~\cite{Michie2013ABM}, such as \textit{goal and planning} and \textit{feedback and monitoring}.

As shown in Figure~\ref{fig:screenshot_logging}, Foodbot utilizes automatic speech recognition and mobile messaging interface, ubiquitous in most smartphone platforms, to replace traditional food-journal user interface. Users simply dictate what they have eaten through simple dialogue turns. Next, Foodbot relies on a large-scale food knowledge graph to ensure sufficient coverage of food items. A simple key innovation of the food knowledge graph is in its data-driven and open-source approach to constructing an extensive database of locally available food items, sourcing from multiple online food delivery services. Together with the dietary self-tracking data, the food knowledge graph also enables Foodbot to provide personalized food recommendation to the users.

Lastly, Foodbot allows users to set specific dietary intake goals following evidence-based dietary guidelines~\cite{Achananuparp2018DoesJEH}, for example, a fish intake goal is at least two servings per week, a water intake goal is about 6-8 glasses per day, etc. After the goals are set, Foodbot will monitor users' progress toward those goals. As needed, JIT notification prompts will be sent to the users to remind them about their progress toward the goals. Users can also directly interact with Foodbot to inquire about their goals and request personalized food recommendations. The goal-tracking dialogue turns following a JIT prompt are shown in Figure~\ref{fig:screenshot_intervention}.



\section{Related Work}
\label{sec:related-work}

Early human-computer interaction (HCI) research has pioneered the applications of virtual agents in the health-related domains. Those studies typically focus on designing persuasive agents with human-like qualities for promoting physical activities~\cite{Bickmore2005TOCHI,Bickmore2011JBI,Albaina2009Pervasive,Schulman2009Persuasive} and healthy eating~\cite{Mazzotta2007IEEE,Bickmore2011JBI} in various populations. Recently, the widespread adoption of the smartphone \& mobile apps and the substantial advances in wearable \& mobile sensing, machine learning, and artificial intelligence (AI) technologies have renewed interests in conversational agents (commonly known as chatbot) as a promising solution to a population-scale and cost-effective delivery of personalized health interventions. Several studies have explored the use of chatbot for monitoring and promoting behavior change in a wide variety of healthcare domains~\cite{Pereira2019UsingHC}. For health promotion, a new wave of data-centric chatbots for physical activities~\cite{Fadhil2019arXiv,Kramer2019InvestigatingIC} and healthy eating~\cite{Fadhil2017AddressingCI,Gabrielli2018SLOWBotL,Fadhil2019arXiv} have recently been explored.

For chatbots designed for supporting dietary monitoring and personalized feedback, our Foodbot is most similar to CoachAI platform~\cite{Fadhil2019arXiv,Fadhil2017AddressingCI}. Both Foodbot and CoachAI employ similar behavior change techniques. However, CoachAI operates as a telemedicine platform, utilizing chatbot as a front-end user interface for monitoring and delivering interventions, thus requiring full supervision from healthcare practitioners. In contrast, all user interactions 
in Foodbot are completely automated, making it more scalable for population-wide interventions. Next, CoachAI uses wellness questionnaires to periodically collect overall dietary patterns from users, whereas Foodbot focuses on a food journal approach to dietary self-monitoring, which captures more granular data. Furthermore, Foodbot incorporates JIT mechanism in monitoring and providing personalized feedback.

The usage of a natural-language user interface for dietary self-tracking in Foodbot is similar to a nutrition dialogue system in ~\cite{Korpusik2019AFoodLS}, where a speech and language understanding model was trained to recognize 975 food concepts, including quantity, brand, and description. Foodbot depends on services provided by Google Assistant and Dialogflow, which were further tuned, to recognize 177,700 food entities from natural-language inputs.
\section{The Foodbot System}
\label{sec:system-overview}

Foodbot is an open-source chatbot built on top of Dialogflow 
platform and accessible through Google Assistant. 
The Foodbot system, shown in Figure~\ref{fig:system}, consists of three main components, namely \textit{conversation engine}, \textit{core services}, and \textit{data store}. Using Google Assistant, a user can interact with Foodbot via voice commands, clicks (on embedded user interface elements, e.g., cards and suggestion chips), and free-text inputs. 

The conversation engine consists of: (1) Dialogflow services for intent detection, entity recognition, and dialog state management; and (2) a webhook server 
to bridge Dialogflow and core services. With built-in automatic speech recognition service, the conversation engine first translates user-interaction events into specific service requests and then generates personalized responses back to the user. Each time the webhook server received messages from Dialogflow, it triggers an intent fulfilment to execute the corresponding core service to complete the user request provided that the information is complete. If the information given by the user is incomplete or ambiguous, the intent fulfilment will trigger response generation to send follow-up clarification questions.

The core services component 
consists of a collection of web services performing main functionalities, i.e., food logging, food search, food recommendation, and goal setting and tracking. The services are categorized into two types: request-based services (depicted as blue boxes in Figure~\ref{fig:system}), and JIT intervention services. The request-based services execute intent fulfilment requests. The JIT intervention relies on a job scheduler to invoke the services. The intervention triggers intent fulfilment to generate a response as notification prompt. Currently, JIT intervention delivers food recommendation and goal-tracking reminder. 

Lastly, the data store component holds food knowledge graph and user data, including food journal data. The user data are stored in MySQL database and food knowledge graph in Elasticsearch.

\subsection{Conversation Engine}
\label{sec:conversation-engine}

The conversation engine is responsible for processing natural-language inputs, determining user intents and corresponding tasks, and managing dialog state. These are handled by three main modules in the Dialogflow platform.
Given an input text (e.g., transcribed voice commands or free-form text messages), the intent detection module infers user intent from the input while, in parallel, the entity extraction module identifies relevant entities and information in the input. We train machine-learning models in Dialogflow by supplying 20-30 examples for each intent. Periodically, we add new examples gathered from past user interactions with Foodbot and retrain the models to improve the detection accuracy.


Next, the dialog state management module keeps track of conversation state between the user and Foodbot and determines follow-up intents based on contexts (e.g., previous responses). The intent fulfillment module acts as a router, bridging the user's natural language inputs and core services. Utilizing dialog state information together with user intent, context, and message, the intent fulfilment module decides whether Foodbot should ask a follow-up question or proceed to execute the user request. Lastly, the response generation module generates responses based on a pre-defined set of templates (i.e., canned responses) manually crafted for specific intents.

\subsection{Core Services}
\label{sec:core-services}

The core services component provides back-end functionalities to the conversation engine. Utilizing food knowledge graph and user data, the core services component executes intent fulfilment requests and JIT intervention. Three main functionalities are offered through the following modules: (1) food logging; (2) personalized food recommendation; and (3) goal-setting and JIT intervention. The core services component also provides a general food item retrieval service based on food name, category, and location.

\textbf{Food Logging.}
The food logging module provides basic services for food journaling. These services are essential to other modules, such as personalized recommendation and goal tracking, as they utilize the user's food log entries as the main input data.

The food logging module receives input messages from intent fulfilment containing food name and meal occasion (i.e., breakfast, lunch, dinner, or snack). The input food name will be matched with existing food entities in the food knowledge graph before a food log entry is recorded. If the exact match cannot be found, the user will be shown a list of top-15 most similar food items and asked to select one from the list. The user can also log water intake through Foodbot. Optionally, the user is able to view and modify past food log entries by accessing their food journal interface. To avoid lapses in recording food and water intakes, Foodbot reminds the user to log their food entries on a daily basis via notification prompts.

\textbf{Personalized Recommendation.}
This module generates personalized food recommendation based on the user's past food intake data. The recommendations are delivered by either normal requests or JIT intervention. That is, users can directly ask Foodbot for food recommendation for a specific meal occasion. Before regular meal time, JIT intervention triggers food recommendation to be pushed to users. For example, Foodbot sends food recommendation for lunch to users at around 11 AM everyday.

Currently, the personalized recommendation module employs a simple rule-based algorithm to suggest food items for each meal occasion based on the user's recent and frequent consumption records. This method tends to yield higher recommendation accuracy than most state-of-the-art recommendation algorithms, but at the expense of novel choices~\cite{Liu2019CharacterizingAP}. When past consumption data are not available (e.g., for new users), the most globally popular items will be used for recommendations instead. To improve the user experience from personalized recommendation, more sophisticated algorithms and contextual information will be integrated into Foodbot in the future.


\textbf{Goal-Setting and JIT Intervention.}
Users can ask Foodbot to keep track of a number of dietary intake goals. By default, the goal-setting module recommends a few intake goals according to evidence-based dietary guidelines~\cite{Achananuparp2018DoesJEH}.
The JIT intervention module identifies windows of opportunity to nudge the users toward their goals and avoids generating excessive prompts, which might lead to notification fatigue. Together, the goal-setting and JIT intervention modules provide \textit{just-in-time}, \textit{personalized}, and \textit{actionable} feedback to users. 


To determine progress toward the food intake goals from the user's food logging data, we utilize food category labels in the food knowledge graph. These labels correspond to basic dietary constituents, i.e., fruit \& vegetable, red \& processed meat, fish, added sugar, nut, and water. A serving amount is determined based on the frequency of relevant keywords in the user's food log entries matching the corresponding labels. For example, if the food journal contains two entries for \textit{apple} and \textit{banana}, the current progress for fruit \& vegetable goal is 2 servings. At the moment, this simple approach is good enough for general behavior monitoring since Foodbot does not currently consider precise nutritional values, e.g., calories, as parts of the dietary goals.

Toward the end of the day, the JIT intervention module will send a notification prompt to the user if the gap between current and target intakes exceeds certain thresholds. Additional conditions may also be applied to different type of goals.
For example, for weekly fish intake goal, Foodbot will remind the user if no fish intake is observed before the dinner time on Wednesday (the middle of the week). At the end of the week, Foodbot will present the user with a weekly progress report to help the user reflect on their past consumption and, if needed, adjust their future behavior. If the user fails to achieve any of the goals, Foodbot will propose a new set of attainable goals.


\subsection{Food Knowledge Graph}
\label{sec:fkg}

Food knowledge graph supports various core services' modules by providing a comprehensive knowledge base of local food items available from restaurants and grocery stores in Singapore (the location in which the target users of Foodbot reside). The food knowledge graph is constructed from an open-source online data about local foods and restaurants. To collect such raw data, we develop our own web crawlers to scrape and aggregate relevant information from the following online sources: (1) Online food delivery services Deliveroo and Foodpanda; (2) Burpple, a popular online social network for food and restaurant reviews; (3) Online grocery shopping site NTUC FairPrice, the largest local supermarket chains.

We believe that these sources contain a wide range of foods and beverages which are commonly consumed in Singapore. Given the initial crawled data, we perform data cleaning (e.g., removing duplicate records, text parsing and normalizing, etc.) and data integration to derive a unique set of food items representing entities in the knowledge graph. In total, there are 177,700 unique entities comprising: (a) 158,211 food items from local restaurants, food courts, and hawker centres in Singapore;  (b) 15,588 packaged food items from grocery stores, and (c) 3,901 generic food items imported from food taxonomy, created by~\cite{Weber2015InsightsFM}. The inclusion of generic food items is to better handle naming variations.

Schematically, the food knowledge graph contains two types of entities representing food items and restaurants. Each type of entities has its own corresponding attributes. For example, food entity’s attributes include name variations, descriptions, prices, and reviews, whereas restaurant entity’s attributes include location, contact number, and cuisine type. We also enrich food entities with food item categories from food taxonomy~\cite{Weber2015InsightsFM}, utilizing the categories for goal tracking service as explained in Section~\ref{sec:core-services}. Two types of linkages define relationships between food-restaurant entities (availability) and food-food entities (similarity).

\section{Conclusion and Future Work}
\label{sec:future-work}

We present an open-source chatbot for healthy eating intervention that runs on the Google Assistant platform. Powered by a large-scale food knowledge graph, Foodbot provides a natural-language user interface to reduce barriers in dietary self-tracking. In addition, it guides users toward setting realistic healthy eating goals according to evidence-based dietary guidelines and nudges them to achieve their goals via JIT personalized actionable feedback. Foodbot can be accessed at \url{https://research.larc.smu.edu.sg/foodbot}. 


For future work, we plan to conduct user studies to evaluate the usability of Foodbot and, subsequently, the effectiveness of Foodbot in healthy lifestyle interventions in the general population. Next, the food knowledge graph can be enriched with ingredients and nutritional information to improve the quality of food recommendation and expand the scope of goal-setting. Lastly, we wish to utilize natural-language generation models to improve response generation, enabling automatic generation of personalized and persuasive responses.

\begin{acks}
This research is supported by the National Research Foundation, Singapore under its International Research Centres in Singapore Funding Initiative. Any opinions, findings and conclusions or recommendations expressed in this material are those of the author(s) and do not reflect the views of National Research Foundation, Singapore.
\end{acks}

\bibliographystyle{ACM-Reference-Format}
\bibliography{references-final}

\end{document}